\documentclass[sigconf]{acmart}

\usepackage{booktabs} 

\usepackage{amsmath,url}
\usepackage{graphicx}
\usepackage{color}
\usepackage{xspace}
\usepackage{siunitx}
\usepackage{caption}
\usepackage{subcaption}
\usepackage{float}
\usepackage{lipsum}
\usepackage{color}
\usepackage{tikz}
\usepackage{pgfmath}

\usetikzlibrary{arrows.meta}

\setcopyright{acmcopyright}

\acmConference{DLfM}{Sep. 2018}{Paris, France}
\acmYear{2018}
\copyrightyear{2018}

\begin{document}
\title{Extended playing techniques:\\
The next milestone in musical instrument recognition}

\author{Vincent Lostanlen}
\orcid{0000-0003-0580-1651}
\affiliation{%
  \institution{New York University}
  \streetaddress{35 W 4th St}
  \city{New York}
  \state{NY, USA}
  \postcode{10014}
}
\email{vincent.lostanlen@nyu.edu}

\author{Joakim And\'{e}n}
\orcid{0000-0002-3377-813X}
\affiliation{%
  \institution{Flatiron Institute}
  \streetaddress{162 5th Ave}
  \city{New York}
  \state{NY, USA}
  \postcode{10010}
}
\email{janden@flatironinstitute.edu}

\author{Mathieu Lagrange}
\affiliation{%
  \institution{\'{E}cole Centrale de Nantes, CNRS}
  \streetaddress{1, rue de la No\"{e}}
  \city{44321 Nantes}
  \state{France}
  \postcode{43017-6221}
}
\email{mathieu.lagrange@cnrs.fr}

\renewcommand{\shortauthors}{V. Lostanlen et al.}

\newcommand\blfootnote[1]{%
  \begingroup
  \renewcommand\thefootnote{}\footnote{#1}%
  \addtocounter{footnote}{-1}%
  \endgroup
}

\begin{abstract}
The expressive variability in producing a musical note conveys information essential to the modeling of orchestration and style.
As such, it plays a crucial role in computer-assisted browsing of massive digital music corpora.
Yet, although the automatic recognition of a musical instrument from the recording of a single ``ordinary'' note is considered a solved problem, automatic identification of instrumental playing technique (IPT) remains largely underdeveloped.
We benchmark machine listening systems for query-by-example browsing among $143$ extended IPTs for $16$ instruments, amounting to $469$ triplets of instrument, mute, and technique. We identify and discuss three necessary conditions for significantly outperforming the traditional mel-frequency cepstral coefficient (MFCC) baseline: the addition of second-order scattering coefficients to account for amplitude modulation, the incorporation of long-range temporal dependencies, and metric learning using large-margin nearest neighbors (LMNN) to reduce intra-class variability.
Evaluating on the Studio On Line (SOL) dataset, we obtain a precision at rank $5$ of $99.7\%$ for instrument recognition (baseline at $89.0\%$) and of $61.0\%$ for IPT recognition (baseline at $44.5\%$).
We interpret this gain through a qualitative assessment of practical usability and visualization using nonlinear dimensionality reduction.
\blfootnote{
The source code to reproduce the experiments of this paper is made available at: \url{https://www.github.com/mathieulagrange/dlfm2018}}
\end{abstract}

%
%



\maketitle

\title{Extended playing techniques: the next milestone in musical instrument recognition}

\newcommand*{\eg}{e.g.\@\xspace}
\newcommand*{\ie}{i.e.\@\xspace}
\newcommand*{\resp}{resp.\@\xspace}
\newcommand*{\vs}{vs.\@\xspace}

\newcommand{\vl}[1]{\textcolor{orange}{Vincent: #1}}
\newcommand{\ja}[1]{\textcolor{magenta}{Joakim: #1}}
\renewcommand{\ml}[1]{\textcolor{blue}{Mathieu: #1}}


\newcommand{\scal}[1]{%
\tikzset{>={Stealth[width=1.5mm,length=2mm]}}%
\begin{tikzpicture}%
    \node[anchor=south west,inner sep=0] at (0, 0) {\includegraphics[width=\scalwidth]{#1}};%
    \draw[->] (0, 0) -- (\scalwidth + \arrowextra, 0);%
    \node[anchor=west] at (\scalwidth + \arrowextra, 0) {\footnotesize $t$};%
    \draw[->] (0, 0) -- (0, \scalheight + \arrowextra);%
    \node[anchor=east] at (0, \scalheight) {\footnotesize $\lambda_1$};%
\end{tikzpicture}%
}


\section{Introduction}

\begin{figure}
        \newcommand{\scalwidth}{35.0mm}
        \newcommand{\scalheight}{27.7mm}
        \newcommand{\arrowextra}{2mm}

        \begin{subfigure}{0.25\textwidth}
                \centering
                \scal{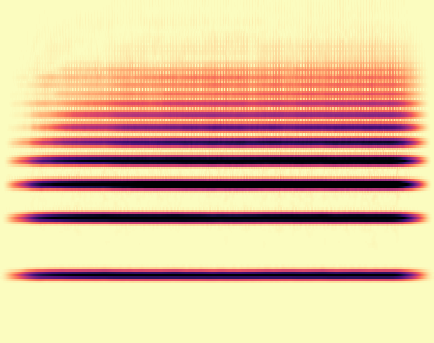}
                \caption{Trumpet note (\emph{ordinario}).}
                \label{fig:TpC-ord-G4-mf_withaxes}
        \end{subfigure}%
        \begin{subfigure}{0.25\textwidth}
                \centering
                \scal{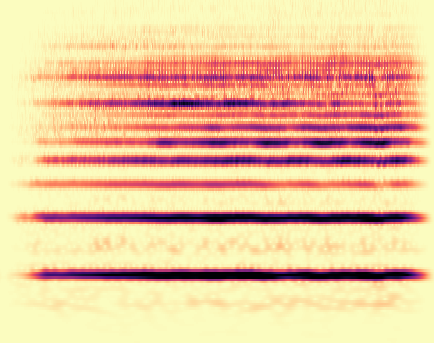}
                \caption{Instrument (violin).}
                \label{fig:Vn-ord-G4-mf-4c_withaxes}
        \end{subfigure}

        \vspace{3mm}

        \begin{subfigure}{0.25\textwidth}
                \centering
                \scal{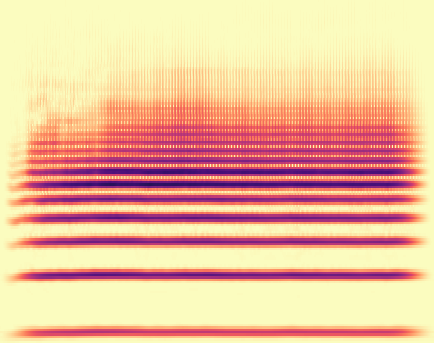}
                \caption{Pitch (G3).}
                \label{fig:TpC-ord-G3-mf_withaxes}
        \end{subfigure}%
        \begin{subfigure}{0.25\textwidth}
                \centering
                \scal{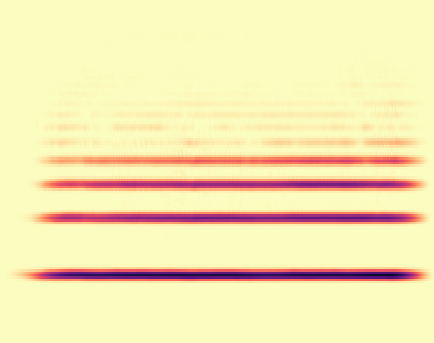}
                \caption{Intensity (\emph{pianissimo}).}
                \label{fig:TpC-ord-G4-pp_withaxes}
        \end{subfigure}

        \vspace{3mm}

        \begin{subfigure}{0.25\textwidth}
                \centering
                \scal{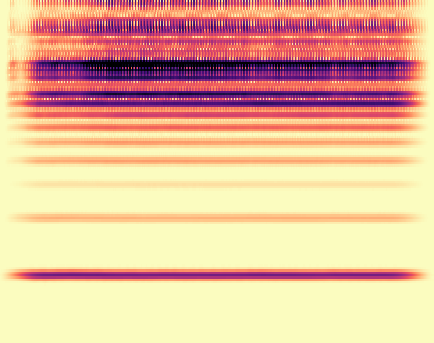}
                \caption{Mute (\emph{harmon}).}
                \label{fig:TpC+H-ord-G4-mf_withaxes}
        \end{subfigure}%
        \begin{subfigure}{0.25\textwidth}
                \centering
                \scal{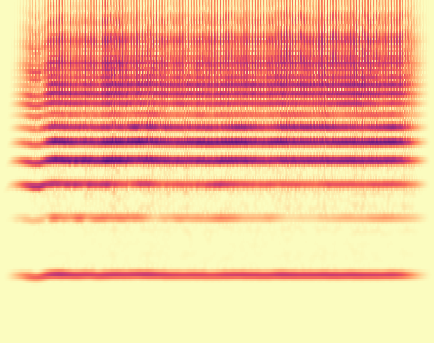}
                \caption{Tone quality (brassy).}
                \label{fig:TpC-brassy-G4-mf_withaxes}
        \end{subfigure}%

        \vspace{3mm}

        \begin{subfigure}{0.25\textwidth}
                \centering
                \scal{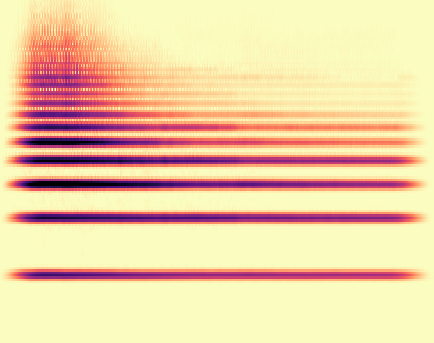}
                \caption{Attack (\emph{sfzorzando}).}
                \label{fig:TpC-sfz-G4-fp_withaxes}
        \end{subfigure}%
        \begin{subfigure}{0.25\textwidth}
                \centering
                \scal{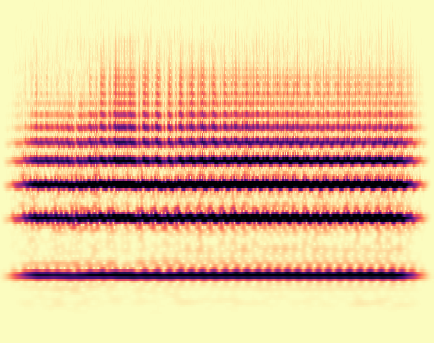}
                \caption{Tonguing (\emph{flatterzunge}).}
                \label{fig:TpC-flatt-G4-mf_withaxes}
        \end{subfigure}

        \vspace{3mm}

        \begin{subfigure}{0.25\textwidth}
                \centering
                \scal{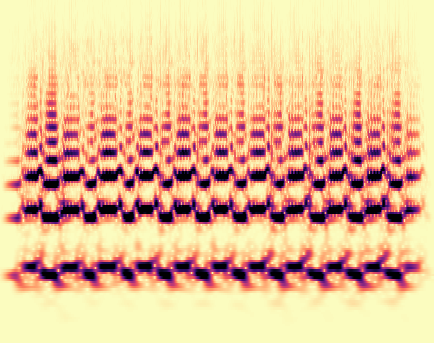}
                \caption{Articulation (\emph{trill}).}
                \label{fig:TpC-trill-maj2-G4-mf_withaxes}
        \end{subfigure}%
        \begin{subfigure}{0.25\textwidth}
                \centering
                \scal{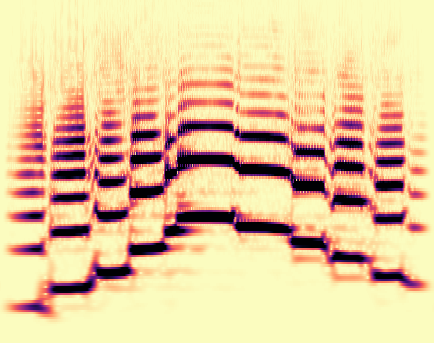}
                \caption{Phrasing (\emph{d\'{e}tach\'{e}}).}
                \label{fig:TpC+voc-harms-G4-mf_withaxes}
        \end{subfigure}%
        \caption{Ten factors of variations of a musical note: pitch (\ref{fig:TpC-ord-G3-mf_withaxes}), intensity (\ref{fig:TpC-ord-G4-pp_withaxes}), tone quality (\ref{fig:TpC-brassy-G4-mf_withaxes}), attack (\ref{fig:TpC-sfz-G4-fp_withaxes}), tonguing (\ref{fig:TpC-flatt-G4-mf_withaxes}), articulation (\ref{fig:TpC-trill-maj2-G4-mf_withaxes}), mute (\ref{fig:TpC+H-ord-G4-mf_withaxes}), phrasing (\ref{fig:TpC+voc-harms-G4-mf_withaxes}), and instrument (\ref{fig:Vn-ord-G4-mf-4c_withaxes}).}\label{fig:trumpet-variations}
\end{figure}

The gradual diversification of the timbral palette in Western classical music since the dawn of the 20th century is reflected in five concurrent trends:
the addition of new instruments to the symphonic instrumentarium, either by technological inventions (\eg theremin) or importation from non-Western musical cultures (\eg marimba) \cite[epilogue]{sachs2012book};
the creation of novel instrumental associations, as epitomized by \emph{Klangfarbenmelodie} \cite[chapter 22]{schoenberg2010book};
the temporary alteration of resonant properties through mutes and other ``preparations'' \cite{dianova2007phd};
a more systematic usage of extended instrumental techniques, such as artificial harmonics, \emph{col legno batutto}, or flutter tonguing \cite[chapter 11]{kostka2016book};
and the resort to electronics and digital audio effects \cite{zolzer2011dafx}.
The first of these trends has somewhat stalled.
To this day, most Western composers rely on an acoustic instrumentarium that is only marginally different from the one that was available in the Late Romantic period.
Nevertheless, the remaining trends in timbral diversification have been adopted on a massive scale in post-war contemporary music.
In particular, an increased concern for the concept of musical gesture \cite{godoy2009book} has liberated many unconventional instrumental techniques from their figurativistic connotations, thus making the so-called ``ordinary'' playing style merely one of many compositional -- and improvisational -- options.

Far from being exclusive to contemporary music, extended playing techniques are also commonly found in oral tradition; in some cases, they even stand out as a distinctive component of musical style.
Four well-known examples are
the snap pizzicato (``slap'') of the upright bass in rockabilly,
the growl of the tenor saxophone in rock'n'roll,
the shuffle stroke of the violin (``fiddle'') in Irish folklore,
and the glissando of the clarinet in Klezmer music.
Consequently, the organology (the instrumental \emph{what?}) of a recording, as opposed to its chironomics (the gestural \emph{how?}), is a poor organizing principle for browsing and recommendation in large music databases.

Yet, past research in music information retrieval (MIR), and especially in machine listening, rarely acknowledges the benefits of integrating the influence of performer gesture into a coherent taxonomy of musical instrument sounds.
Instead, gesture is often framed as a spurious form of intra-class variability between instruments without delving into its interdependencies with pitch and intensity.
In other works, it is conversely used as a probe for the acoustical study of a given instrument without emphasis on the broader picture of orchestral diversity.

One major cause of this gap in research is the difficulty of collecting and annotating data for contemporary instrumental techniques.
Fortunately, this obstacle has recently been overcome, owing to the creation of databases of instrumental samples for music orchestration in spectral music \cite{maresz2013cmr}.
In this work, we capitalize on the availability of this data to formulate a new line of research in MIR, namely the joint retrieval of organological (``\emph{what} instrument is being played in this recording?'') and chironomical information (``\emph{how} is the musician producing sound?''), while remaining invariant to other factors of variability deliberately regarded as contextual.
These include at what pitch and intensity the music was recorded, but also where, when, why, by whom, and for whom it was created.

Figure \ref{fig:TpC-ord-G4-mf_withaxes} shows the constant-$Q$ wavelet scalogram (\ie{} the complex modulus of the constant-$Q$ wavelet transform) of a trumpet musical note, as played with an ordinary technique.
Unlike most existing publications on instrument classification (\eg{} \ref{fig:TpC-ord-G4-mf_withaxes} \vs{} \ref{fig:Vn-ord-G4-mf-4c_withaxes}), which exclusively focus on intra-class variability due to pitch (Figure \ref{fig:TpC-ord-G3-mf_withaxes}) and intensity (Figure \ref{fig:TpC-ord-G4-pp_withaxes}), and mute (\ref{fig:TpC+H-ord-G4-mf_withaxes}), this work aims to also account for the presence of instrumental playing techniques (IPTs), such as changes in tone quality (Figure \ref{fig:TpC-brassy-G4-mf_withaxes}), attack (Figure \ref{fig:TpC-sfz-G4-fp_withaxes}), tonguing (Figure \ref{fig:TpC-flatt-G4-mf_withaxes}), and articulation (Figure \ref{fig:TpC-trill-maj2-G4-mf_withaxes}).
These factors are considered either as intra-class variability, for the instrument recognition task, or as inter-class variability, for the IPT recognition task.
The analysis of IPTs whose definition involves more than a single musical event, such as phrasing (Figure \ref{fig:TpC+voc-harms-G4-mf_withaxes}), is beyond the scope of this paper.

Section 2 reviews the existing literature on the topic.
Section 3 defines taxonomies of instruments and gestures from which the IPT classification task is derived.
Section 4 describes how two topics in machine listening, namely characterization of amplitude modulation and incorporation of supervised metric learning, are relevant to address this task.
Section 5 reports the results from an IPT classification benchmark on the Studio On Line (SOL) dataset.

\section{Related work}
This section reviews recent MIR literature on the audio analysis of IPTs with a focus on the datasets available for the various classification tasks considered.

\subsection{Isolated note instrument classification}
The earliest works on musical instrument recognition restricted their scope to individual notes played with an ordinary technique, eliminating most factors of intra-class variability due to the performer \cite{martin1998asa,brown1999jasa,eronen2000icassp,herrera2003jnmr,wieczorkowska2003jiis,kaminskyj2005jiis,benetos2006icassp}.
These results were obtained on datasets such as MUMS \cite{opolko1989dataset}, MIS,\footnote{\url{http://theremin.music.uiowa.edu/MIS.html}} RWC \cite{goto2003ismir}, and samples from the Philharmonia Orchestra.\footnote{\url{http://www.philharmonia.co.uk/explore/sound_samples}}
This line of work culminated with the development of a support vector machine classifier trained on spectrotemporal receptive fields (STRF), which are idealized computational models of neurophysiological responses in the central auditory system \cite{chi2005jasa}.
Not only did this classifier attain a near-perfect mean accuracy of $98.7\%$ on the RWC dataset, but the confusion matrix of its predictions was close to that human listeners \cite{patil2012plos}.
Therefore, supervised classification of instruments from recordings of ordinary notes could arguably be considered a solved problem; we refer to \cite{bhalke2016jiis} for a recent review of the state of the art.

\subsection{Solo instrument classification}
A straightforward extension of the problem above is the classification of solo phrases, encompassing some variability in melody \cite{krishna2004icassp}, for which the accuracy of STRF models is around $80\%$ \cite{patil2015eurasip}.
Since the Western tradition of solo music is essentially limited to a narrow range of instruments (\eg{} piano, classical guitar, violin) and genres (sonatas, contemporary, free jazz, folk), datasets of solo phrases, such as solosDb \cite{joder2009taslp}, are exposed to strong biases.
This issue is partially mitigated by the recent surge of multitrack datasets, such as MedleyDB \cite{bittner2014ismir}, which has spurred a renewed interest in single-label instrument classification \cite{yip2017ismir}.
In addition, the cross-collection evaluation methodology \cite{livshin2003ismir} reduces the risk of overfitting caused by the relative homogeneity of artists and recording conditions in these small datasets \cite{bogdanov2016ismir}.
To date, the best classifiers of solo recordings are the joint time-frequency scattering transform \cite{anden2018tsp} and the spiral convolutional network \cite{lostanlen2016ismir} trained on the Medley-solos-DB dataset \cite{lostanlen2018msdb}, \ie{}, a cross-collection dataset which aggregates MedleyDB and solosDb following the procedure of \cite{donnelly2015icdmw}.
We refer to \cite{han2017taslp} for a recent review of the state of the art.

\subsection{Multilabel classification in polyphonic mixtures}
Because most publicly released musical recordings are polyphonic, the generic formulation of instrument recognition as a multilabel classification task is the most relevant for many end-user applications \cite{martins2007ismir,burred2009icassp}.
However, it suffers from two methodological caveats.
First, polyphonic instrumentation is not independent from other attributes, such as geographical origin, genre, or key.
Second, the inter-rater agreement decreases with the number of overlapping sources \cite[chapter 6]{fuhrmann2012phd}.
These problems are all the more troublesome since there is currently no annotated dataset of polyphonic recordings diverse enough to be devoid of artist bias.
The Open-MIC initiative, from the newly created Community for Open and Sustainable Music and Information Research (COSMIR), is working to mitigate these issues in the near future \cite{mcfee2016ismir}.
We refer to \cite{humphrey2018ismir} for a recent review of the state of the art.

\subsection{Solo playing technique classification}
Finally, there is a growing interest for studying the role of the performer in musical acoustics, from the perspective of both sound production and perception.
Apart from its interest in audio signal processing, this topic is connected to other disciplines, such as biomechanics and gestural interfaces \cite{metcalf2014frontiers}.
The majority of the literature focuses on the range of IPTs afforded by a single instrument.
Recent examples include clarinet \cite{loureiro2004ismir}, percussion \cite{tindale2004ismir}, piano \cite{bernays2013smc}, guitar \cite{foulon2013cmmr,su2014ismir,chen2015ismir}, violin \cite{young2008nime}, 
and erhu \cite{yang2014fma}.
Some publications frame timbral similarity in a polyphonic setting, yet do so according to a purely perceptual definition of timbre -- with continuous attributes such as brightness, warmth, dullness, roughness, and so forth -- without connecting these attributes to the discrete latent space of IPTs (\ie{}, through a finite set of instructions, readily interpretable by the performer) \cite{antoine2018isma}.
We refer to \cite{leman2017chapter} for a recent review of the state of the art.

In the following, we define the task of retrieving musical timbre parameters across a range of instruments found in the symphonic orchestra.
These parameters are explicitly defined in terms of sound production rather than by means of perceptual definitions.

\section{Tasks}
In this section, we define a taxonomy of musical instruments and another for musical gestures, which are then used for defining the instrument and IPT query-by-example tasks.
We also describe the dataset of instrument samples used in our benchmark.

\subsection{Taxonomies}

\begin{figure}[t!]
\centering
\includegraphics[width=\linewidth]{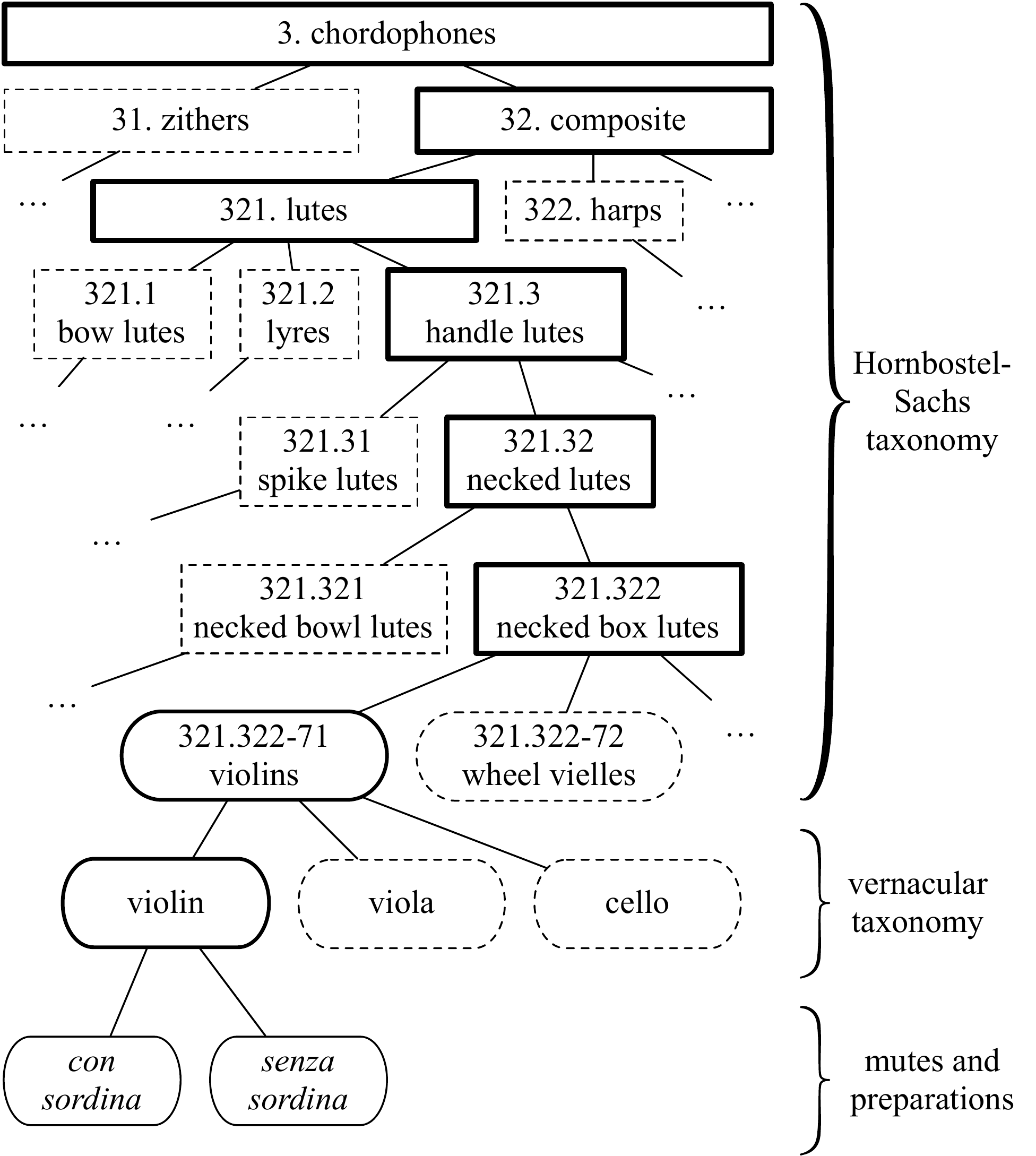}
\caption{Taxonomy of musical instruments.}
\label{fig:instrument-dendrogram}
\end{figure}

The Hornbostel-Sachs taxonomy (H-S) organizes musical instruments only according to their physical characteristics and purposefully ignores sociohistorical background \cite{montagu2009muzyka}.
Since it offers an unequivocal way of describing any acoustic instrument without any prior knowledge of its applicable IPTs, it serves as a \emph{lingua franca} in ethnomusicology and museology, especially for ancient or rare instruments which may lack available informants.
The classification of the violin in H-S (321.322-71), as depicted in Figure \ref{fig:instrument-dendrogram}, additionally encompasses the viola and the cello.
The reason is that these three instruments possess a common morphology.
Indeed, both violin and viola are usually played under the jaw and the cello is held between the knees, these differences in performer posture are ignored by the H-S classification.
Accounting for these differences begs to refine H-S by means a vernacular taxonomy.
Most instrument taxonomies in music signal processing, including MedleyDB \cite{bittner2014ismir} and AudioSet \cite{gemmeke2017icassp}, adopt the vernacular level rather than conflating all instruments belonging to the same H-S class.
A further refinement includes potential alterations to the manufactured instrument -- permanent or temporary, at the time scale one or several notes -- that affect its resonant properties, \eg{}, mutes and other preparations \cite{dianova2007phd}.
The only node in the MedleyDB taxonomy which reaches this level of granularity is \emph{tack piano} \cite{bittner2014ismir} .
In this work, we will not consider variability due to the presence of mutes as discriminative, both for musical instruments and IPTs.

\begin{figure}
\centering
\includegraphics[width=\linewidth]{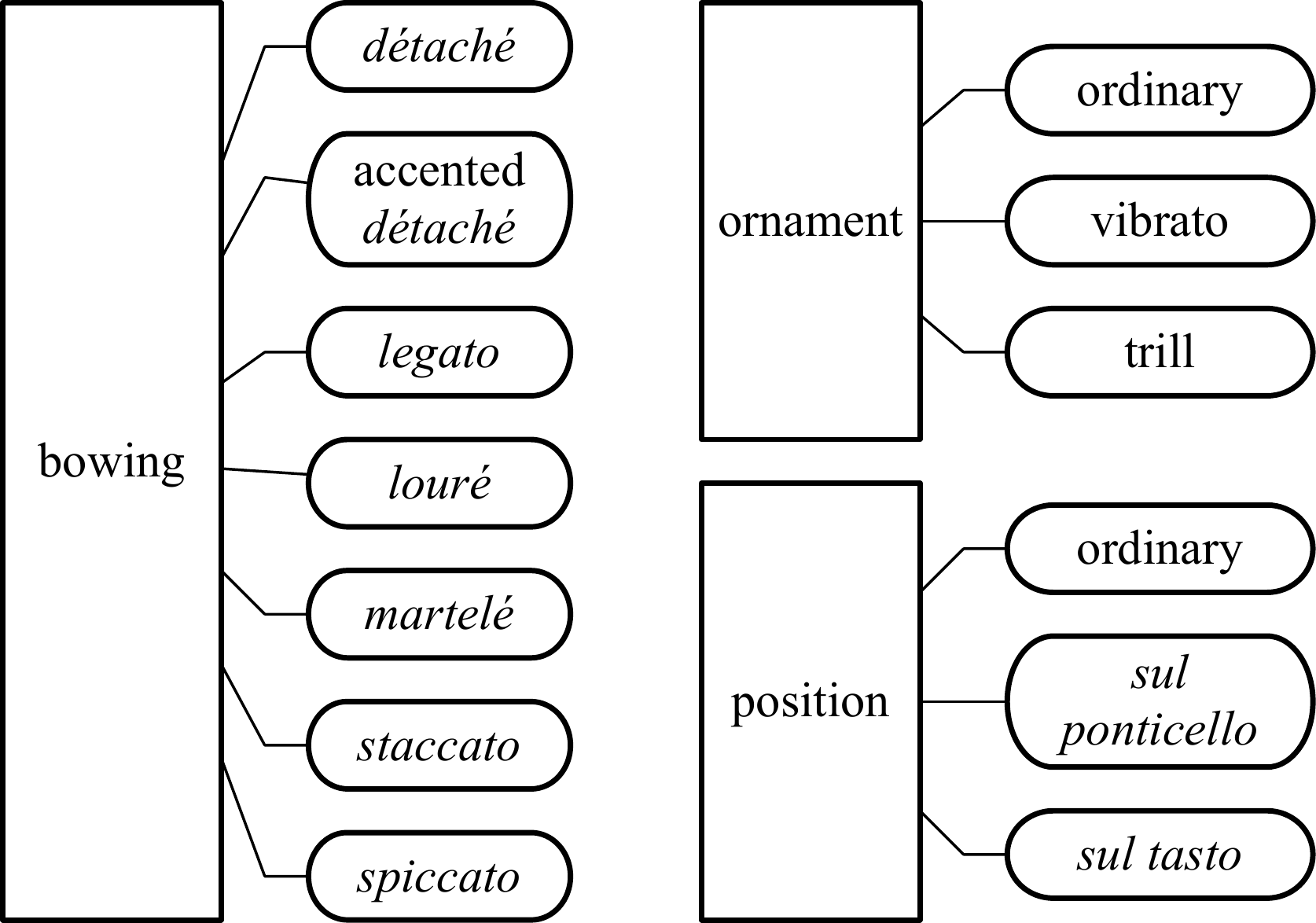}
\caption{Namespaces of violin playing techniques.}
\label{fig:technique-dendrogram}
\end{figure}

Unlike musical instruments, which are amenable to a hierarchical taxonomy of resonating objects, IPTs result from a complex synchronization between multiple gestures, potentially involving both hands, arms, diaphragm, vocal tract, and sometimes the whole body.
As a result, they cannot be trivially incorporated into H-S, or indeed any tree-like structure \cite{kolozali2011ismir}.
Instead, an IPT is described by a finite collection of categories, each belonging to a different ``name\-space.''
Figure \ref{fig:technique-dendrogram} illustrates such namespaces for the case of the violin.
It therefore appears that, rather than aiming for a mere increase in granularity with respect to H-S, a coherent research program around extended playing techniques should formulate them as belonging to a meronomy, \ie{}, a modular entanglement of part-whole relationships, in the fashion of the Visipedia initiative in computer vision \cite{belongie2015pattern}.
In recent years, some works have attempted to lay the foundations of such a modular approach, with the aim of making H-S relevant to contemporary music creation \cite{magnusson2017jnmr,weisser2011ytm}.
However, such considerations are still in large part speculative and offer no definitive procedure for evaluating, let alone training, information retrieval systems.

\begin{figure}
\centering
\includegraphics[width=0.98\linewidth]{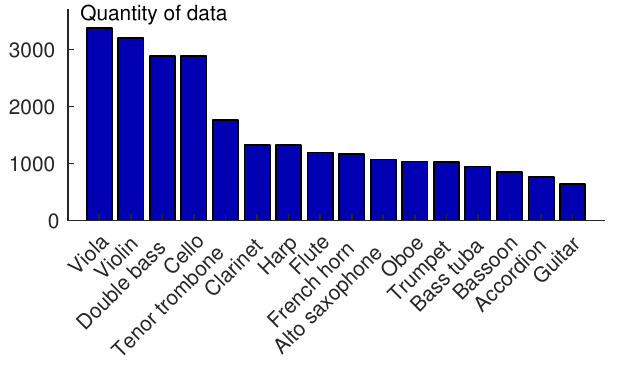}
\caption{Instruments in the SOL dataset.}
\label{fig:instrument-histogram}
\end{figure}
\begin{figure}
\centering
\begin{tikzpicture}
\def\iptarray{{"ordinario","non-vibrato","tremolo","flatterzunge","sforzando","crescendo","note-lasting","pizzicato-l-vib","glissando","decrescendo","pizzicato-secco","staccato","crescendo-to-decrescendo","ordinario-1q","trill-minor-second-up","trill-major-second-up","sul-ponticello","pizzicato-bartok","sul-tasto","sul-ponticello-tremolo","multiphonics","sul-tasto-tremolo","col-legno-battuto","col-legno-tratto","harmonic-fingering","bisbigliando","lip-glissando","artificial-harmonic","ordinario-to-flatterzunge","artificial-harmonic-tremolo","flatterzunge-to-ordinario","vibrato","crushed-to-ordinario","ordinario-to-sul-tasto","ordinario-to-crushed","slap-pitched","sul-ponticello-to-ordinario","sul-ponticello-to-sul-tasto","ordinario-to-tremolo","sul-tasto-to-ordinario","tremolo-to-ordinario","near-the-board","sul-tasto-to-sul-ponticello","ordinario-to-sul-ponticello","aeolian-and-ordinario","brassy","backwards","ordinario-high-register","brassy-to-ordinario","natural-harmonics-glissandi"}}%
\node[anchor=north west] at (2, 0) {\includegraphics[height=149mm]{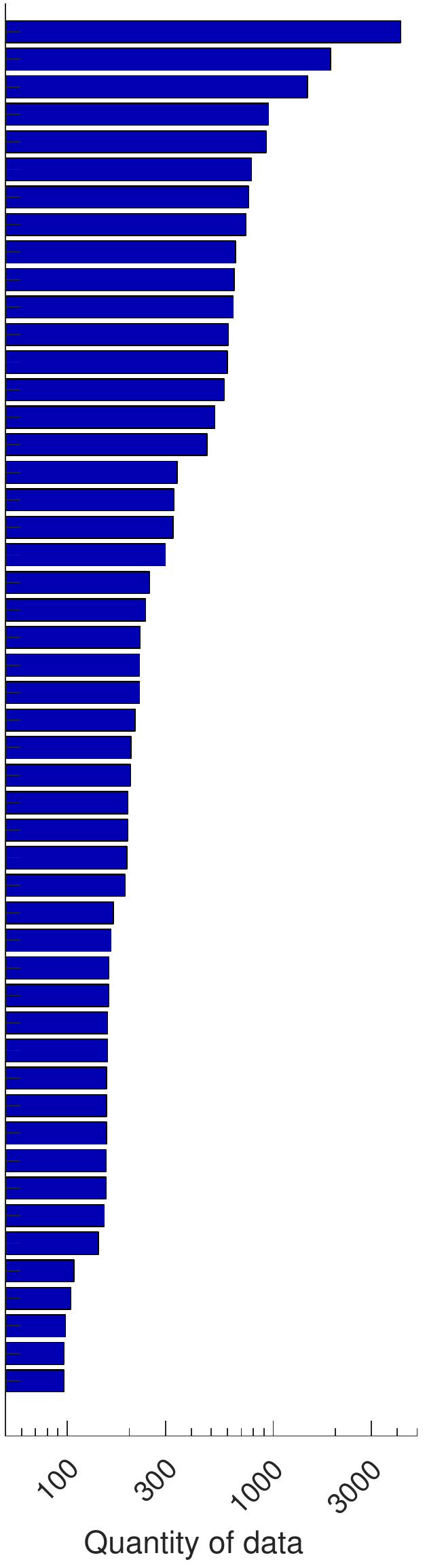}};
\foreach \x in {0,...,49}
    \node[anchor=east] at (2, -0.42 - 0.262*\x) {\footnotesize {\fontfamily{phv}\selectfont\pgfmathparse{\iptarray[\x]}\pgfmathresult}};
\end{tikzpicture}
\caption{The $50$ most common IPTs in the SOL dataset.}
\label{fig:technique-histogram}
\end{figure}

\subsection{Application setting and evaluation} \label{sec:motivation}
In what follows, we adopt a middle ground position between the two aforementioned approaches: neither a top-down multistage classifier (as in a hierarchical taxonomy), nor a caption generator (as in a meronomy), our system is a query-by-example search engine in a large database of isolated notes.
Given a query recording $\boldsymbol{x}(t)$, such a system retrieves a small number $k$ of recordings judged similar to the query.
In our system, we implement this using a $k$-nearest neighbors (k-NN) algorithm.
The nearest neighbor search is not performed in the raw waveform domain of $\boldsymbol{x}(t)$, but in a feature space of translation-invariant, spectrotemporal descriptors.
In what follows, we use mel-frequency cepstral coefficients (MFCCs) as a baseline, which we extend using second-order scattering coefficients \cite{mallat2012cpam,anden2014taslp}.
All features over averaged over the entire recording to create single feature vector.
The baseline $k$-NN algorithm is applied using the standard Euclidean distance in feature space.
To improve performance, we also apply it using a weighted Euclidean distance with a learned weight matrix.

In the context of music creation, the query $\boldsymbol{x}(t)$ may be an instrumental or vocal sketch, a sound event recorded from the environment, a computer-generated waveform, or any mixture of the above \cite{maresz2013cmr}.
Upon inspecting the recordings returned by the search engine, the composer may decide to retain one of the retrieved notes.
Its attributes (pitch, intensity, and playing technique) are then readily available for inclusion in the musical score.

Faithfully evaluating such a system is a difficult procedure, and ultimately depends on its practical usability as judged by the composer.
Nevertheless, a useful quantitative metric for this task is the precision at $k$ (P@$k$) of the test set with respect to the training set, either under an instrument taxonomy and an IPT taxonomy.
This metric is defined as the proportion of ``correct'' recordings returned for a given query, averaged over all queries in the test set.
For our purposes, a returned recording is correct if it is of the same class as the query for a specific taxonomy.
In all subsequent experiments, we report P@$k$ for the number of retrieved items $k=5$.

\subsection{Studio On Line dataset (SOL)}
The Studio On Line dataset (SOL) was recorded at IRCAM in 2002 and is freely downloadable as part of the Orchids software for computer-assisted orchestration.\footnote{\url{http://forumnet.ircam.fr/product/orchids-en/}}
It comprises 16 musical instruments playing $25444$ isolated notes in total.
The distribution of these notes, shown in Figure \ref{fig:instrument-histogram}, spans the full combinatorial diversity of intensities, pitches, preparations (\ie{}, mutes), and all applicable playing techniques.
The distribution of playing techniques is unbalanced as seen in Figure \ref{fig:technique-histogram}.
This is because some playing techniques are shared between many instruments (\eg{}, \textit{tremolo}) whereas other are instrument-specific (\eg{}, \textit{xylophonic}, which is specific to the harp).
The SOL dataset has $143$ IPTs in total, and $469$ applicable instrument-mute-technique triplets.
As such, the dataset has considerable intra-class variability under both the instrument and IPTs taxonomies.



\section{Methods}

In this section, we describe the scattering transform used to capture amplitude modulation structure and supervised metric learning which constructs a similarity measure suited for our query-by-example task.

\subsection{Scattering transform} 
The scattering transform is a cascade of constant-Q wavelet transforms alternated with modulus operators \cite{mallat2012cpam,anden2014taslp}.
Given a signal $\boldsymbol{x} (t)$, its first layer outputs the first-order scattering coefficients $\mathbf{S}_1 \boldsymbol{x} (\lambda_1, t)$, which captures the intensity of $\boldsymbol{x} (t)$ at frequency $\lambda_1$.
Its frequency resolution is logarithmic in $\lambda_1$ and is sampled using $Q_1 = 12$ bins per octave.
The second layer of the cascade yields the second-order scattering coefficients $\mathbf{S}_2 \boldsymbol{x} (\lambda_1, \lambda_2, t)$, which extract amplitude modulation at frequency $\lambda_2$ in the subband of $\boldsymbol{x} (t)$ at frequency $\lambda_1$.
Both first- and second-order coefficients are averaged in time over the whole signal.
The modulation frequencies $\lambda_2$ are logarithmically spaced with $Q_2 = 1$ bin per octave.
In the following, we denote by $\mathbf{S}\boldsymbol{x}(\lambda, t)$ the concatenation of all scattering coefficients, where $\lambda$ corresponds to either a single $\lambda_1$ for first-order coefficients or a pair $(\lambda_1,\lambda_2)$ for second-order coefficients.

The first-order scattering coefficients are equivalent to the mel-frequency spectrogram which forms a basis for MFCCs \cite{anden2014taslp}.
Second-order coefficients, on the other hand, characterize common non-stationary structures in sound production, such as tremolo, vibrato, and dissonance \cite[section 4]{anden2012dafx}.
As a result, these coefficients are better suited to model extended IPTs.
We refer to \cite{anden2014taslp} an introduction on scattering transforms for audio signals and to \cite[sections 3.2 and 4.5]{lostanlen2017phd} for a discussion on its application to musical instrument classification in solo recordings and its connections to STRFs.

To match a decibel-like perception of loudness, we apply the adaptive, quasi-logarithmic compression
\begin{equation}
\widetilde{\mathbf{S}} \boldsymbol{x}_i(\lambda, t) =
\log \left(
1 + \dfrac{\mathbf{S}\boldsymbol{x}_i(\lambda, t)}{\varepsilon \times \boldsymbol{\mu}(\lambda)}
\right)
\label{eq:log-scattering}
\end{equation}
where $\varepsilon = 10^{-3}$ and $\boldsymbol{\mu}(\lambda)$ is the median of $\mathbf{S}\boldsymbol{x}_i (\lambda, t)$ across $t$ and $i$.

\subsection{Metric learning} 
Linear metric learning algorithms construct a matrix $\mathbf{L}$ such that the weighted distance
\begin{equation}
\mathrm{D}_\mathbf{L}(\boldsymbol{x}_i, \boldsymbol{x}_j) = \Vert \mathbf{L} (\widetilde{\mathbf{S}}\boldsymbol{x}_i-\widetilde{\mathbf{S}}\boldsymbol{x}_j) \Vert_2
\end{equation}
between all pairs of samples $(\boldsymbol{x}_i, \boldsymbol{x}_j)$ optimizes some objective function.
We refer to \cite{bellet2013survey} for a review of the state of the art.
In the following, we shall consider the large-margin nearest neighbors (LMNN) algorithm.
It attempts to construct $L$ such that for every signal $\boldsymbol{x}_i (t)$ the distance $\mathrm{D}_\mathbf{L}(\boldsymbol{x}_i, \boldsymbol{x}_j)$ to $\boldsymbol{x}_j (t)$, one of its $k$ nearest neighbors, is small if $\boldsymbol{x}_i (t)$ and $\boldsymbol{x}_j (t)$ belong to the same class and large otherwise.
The matrix $\mathbf{L}$ is obtained by applying the special-purpose solver of \cite[appendix A]{weinberger2009jmlr}.
In subsequent experiments, disabling LMNN is equivalent to setting $\mathbf{L}$ to the identity matrix, which yields the standard Euclidean distance on the scattering coefficients $\widetilde{\mathbf{S}} \boldsymbol{x} (\lambda, t)$.

Compared to a class-wise generative model, such a Gaussian mixture model, a global linear model ensures some robustness to minor alterations of the taxonomy.
Indeed, the same learned metric can be applied to similarity measures in related taxonomies without retraining.
This stability is important in the context of IPT, where one performer's $\emph{slide}$ is another's $\emph{glissando}$.
A major drawback of LMNN is its dependency on the standard Euclidean distance for determining nearest neighbors \cite{mcfee2010icml}.
However, this is alleviated for scattering coefficients, since the scattering transform $\mathbf{S} \boldsymbol{x} (t, \lambda)$ is Lipschitz continuous to elastic deformation in the signal $\boldsymbol{x} (t)$ \cite[Theorem 2.16]{mallat2012cpam}.
In other words, the Euclidean distance between the scattering transform of $\boldsymbol{x} (t)$ and a deformed version of the same signal is bounded by the extent of that deformation.

\section{Experimental evaluation} \label{sec:exp}
In this section, we study a query-by-example browsing system for the SOL dataset based on nearest neighbors.
We discuss how the performance of the system is affected by the choice of feature (MFCCs or scattering transforms) and distance (Euclidean or LMNN), both quantitatively and qualitatively.
Finally, we visualize the two feature spaces using nonlinear dimensionality reduction.

\subsection{Instrument recognition}

In the task of instrument recognition, we provide a query $\boldsymbol{x} (t)$ and the system retrieves $k$ recordings $\boldsymbol{x}_1 (t), \ldots, \boldsymbol{x}_k (t)$.
We consider a retrieved recording to be relevant to the query if it corresponds to the same instrument, regardless of pitch, intensity, mute, and IPT. We therefore apply the LMNN with instruments as class labels. This lets us compute the precision at rank $5$ (P@$5$) for a system by counting the number of relevant recordings for each query.

We compare scattering features to a baseline of MFCCs, defined as the $13$ lowest coefficients of the discrete cosine transform (DCT) applied to the logarithm of the $40$-band mel-frequency spectrum.
For the scattering transform, we vary the maximum time scale $T$ of amplitude modulation from \SI{25}{\milli\second} to \SI{1}{\second}.
In the case of the MFCCs, $T=\SI{25}{\milli\second}$ corresponds to the inverse of the lowest audible frequency ($T^{-1}=\SI{40}{\Hz}$).
Therefore, increasing the frame duration beyond this scale has little effect since no useful frequency information would be obtained.

\begin{figure}
\includegraphics[width=\linewidth,keepaspectratio]{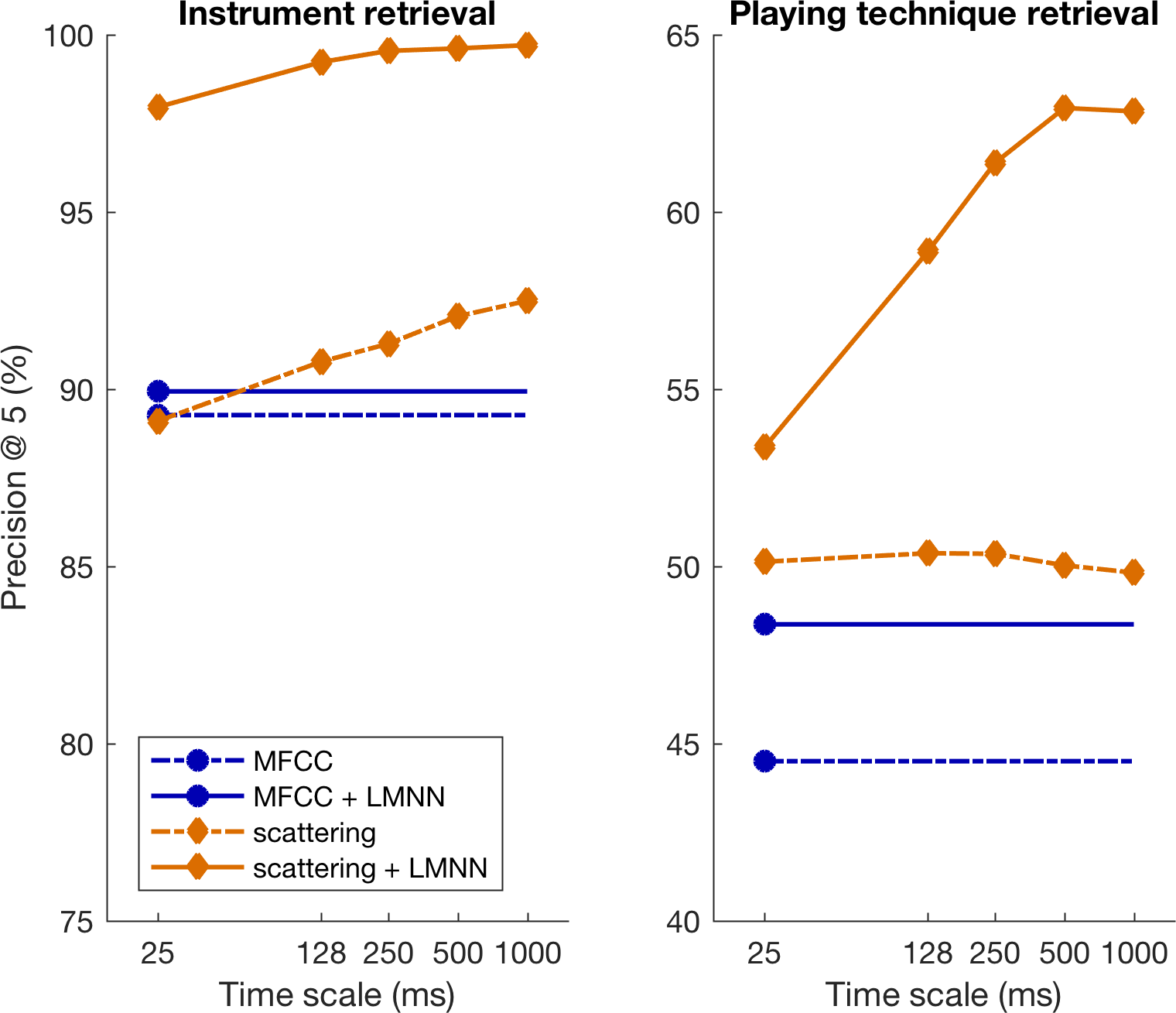}
\caption{Summary of results on the SOL dataset.}
\label{fig:results}
\end{figure}

The left column of Figure \ref{fig:results} summarizes our results.
MFCCs reach a relatively high P@$5$ of $89\%$.
Keeping all $40$ DCT coefficients rather than the lowest $13$ brings P@$5$ down to $84\%$, because the DCT coefficients are most affected by spurious factors of intra-class variability, such as pitch and spectral flatness \cite[subsection 2.3.3]{lostanlen2017phd}.

At the smallest time scale $T=\SI{25}{\milli\second}$, the scattering transform reaches a P@$5$ of $89\%$, thus matching the performance of the MFCCs.
This is expected since there is little amplitude modulation below this scale, corresponding to $\lambda_2$ over $\SI{40}{\Hz}$, so the scattering transform is dominated by the first order, which is equivalent to MFCCs \cite{anden2014taslp}.
Moreover, disabling median renormalization degrades P@$5$ down to $84\%$, while disabling logarithmic compression altogether degrades it to $76\%$.
This is consistent with \cite{lostanlen2018eurasip}, which applies scattering transform to a query-by-example retrieval task for acoustic scenes.

On one hand, replacing the canonical Euclidean distance by a distance learned by LMNN marginally improves P@$5$ for the MFCC baseline, from $89.3\%$ to $90.0\%$.
Applying LMNN to scattering features, on the other hand, significantly improves their performance with respect to the Euclidean distance, from $89.1\%$ to $98.0\%$.

The dimensionality of scattering coefficients is significantly higher than that of MFCCs, which only consists of $13$ coefficients.
A concern is therefore that the higher dimensionality of the scattering coefficients may result in overfitting of the metric learning algorithm, artificially inflating its performance.
To address this, we supplement the averaged MFCCs by higher-order summary statistics.
In addition the $13$ average coefficients, we also compute the average of all polynomial combinations of degree less than three.
The resulting vector is of dimension $494$, comparable to the that of the scattering vector.
This achieves a P@$5$ of $91\%$, that is, slightly above the baseline.
The increased performance of the scattering transform is therefore not likely due overfitting but to its better characterization of multiresolution structure.

Finally, increasing $T$ from $\SI{25}{\milli\second}$ up to $\SI{1}{\second}$ -- \ie{}, including all amplitude modulations between $\SI{1}{\Hz}$ and $\SI{40}{\Hz}$ -- brings LMNN to a near-perfect P@$5$ of $99.7\%$.
Not only does this result confirm that straightforward techniques in audio signal processing (here, wavelet scattering and metric learning) are sufficient to retrieve the instrument from a single ordinary note, it also demonstrates that the results remain satisfactory despite large intra-class variability in terms of pitch, intensity, usage of mutes, and extended IPTs.
In other words, the monophonic recognition of Western instruments is, all things considered, indeed a solved problem.

\subsection{Playing technique recognition}

The situation is different when considering IPT, rather than instrument, as the reference for evaluating the query-by-example system.
In this setting, a retrieved item is considered relevant if and only if it shares the same IPT as the query, regardless of instrument, mute, pitch, or dynamics.
Therefore, we apply the LMNN with IPTs instead of instruments as class labels, yielding a different distance function optimized to distinguish playing techniques.
The right column if Figure \ref{fig:results} summarizes our results.
The MFCC baseline has a low P@$5$ of $44.5\%$, indicating that its coarse description of the short-term spectral envelope is not sufficient to model acoustic similarity in IPT.
Perhaps more surprisingly, we find that optimal performance is only achieved by combining all proposed improvements: log-scattering coefficients with median renormalization, $T=\SI{500}{\milli\second}$, and LMNN.
This yields a P@$5$ of $63.0\%$.
Indeed, an ablation study of that system reveals that, all other things being equal, reducing $T$ to $\SI{25}{\milli\second}$ brings the P@$5$ to $53.3\%$, disabling LMNN reduces it to $50.0\%$, and replacing scattering coefficients by MFCCs yields $48.4\%$.
This result contrasts with the instrument recognition setting: whereas the improvements brought by the three aforementioned modifications are approximately additive in P@$5$ for musical instruments, they interact in a super-additive manner for IPTs.
In particular, it appears that increasing $T$ above \SI{25}{\milli\second} is only beneficial to IPT similarity retrieval if combined with LMNN.

\begin{figure}
        \newcommand{\scalwidth}{28.0mm}
        \newcommand{\scalheight}{22.1mm}
        \newcommand{\arrowextra}{2mm}

        \begin{subfigure}{\linewidth}
                \centering
                \scal{Vn-ord-G4-mf-4c.png}
                \caption*{Query: Violin, \emph{ordinario}, G4, \emph{mf}, on G string.}
                \label{fig:Vn-ord-G4-mf-4c}
        \end{subfigure}%

        \begin{subfigure}{0.20\textwidth}
                \centering
                \scal{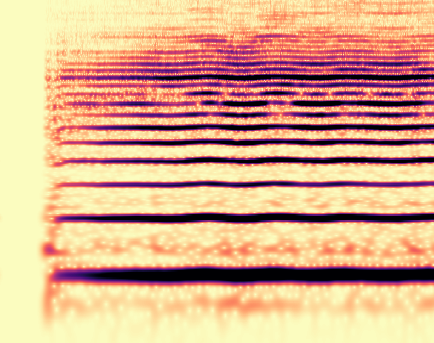}
                \caption*{1: \emph{ff}.}
                \label{fig:Vn-ord-G4-ff-4c}
        \end{subfigure}%
        \begin{subfigure}{0.20\textwidth}
                \centering
                \scal{Vn-ord-G4-ff-4c.png}
                \caption*{1: \emph{ff}.}
                \label{fig:Vn-ord-G4-ff-4c-bis}
        \end{subfigure}%

        \begin{subfigure}{0.20\textwidth}
                \centering
                \scal{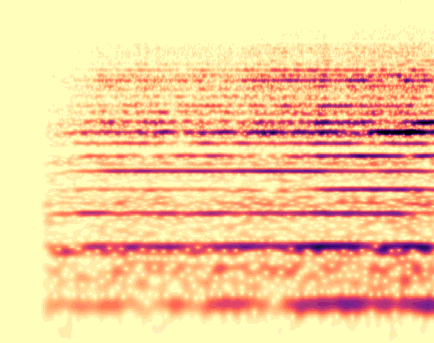}
                \caption*{2: \emph{sul ponticello}, C\#4.}
                \label{fig:Vn-pont-Csh4-mf-4c}
        \end{subfigure}%
        \begin{subfigure}{0.20\textwidth}
                \centering
                \scal{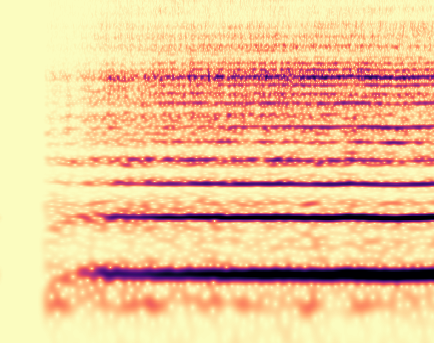}
                \caption*{2: \emph{pp}.}
                \label{fig:Vn-ord-G4-pp-4c}
        \end{subfigure}%

        \begin{subfigure}{0.20\textwidth}
                \centering
                \scal{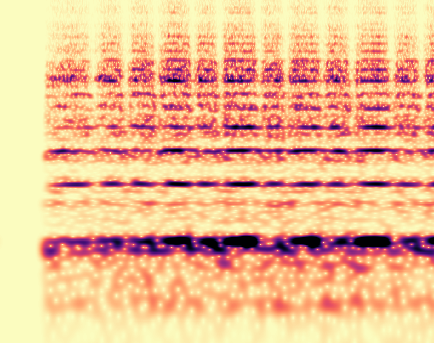}
                \caption*{3: \emph{tremolo}, D5, \emph{pp}.}
                \label{fig:Vn-trem-D5-pp-4c}
        \end{subfigure}%
        \begin{subfigure}{0.20\textwidth}
                \centering
                \scal{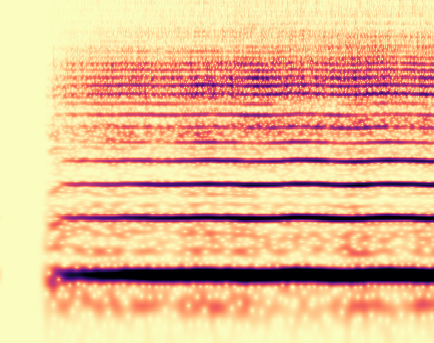}
                \caption*{3: \emph{sordina}.}
                \label{fig:Vn+S-ord-G4-mf-4c}
        \end{subfigure}%

        \begin{subfigure}{0.20\textwidth}
                \centering
                \scal{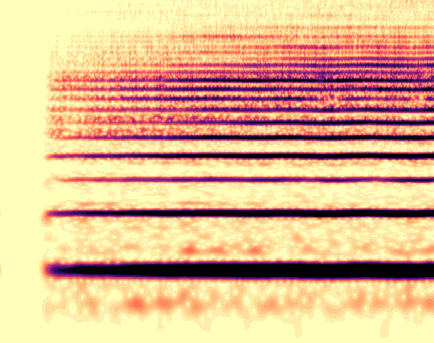}
                \caption*{4: G\#4.}
                \label{fig:Vn-ord-Gsh4-mf-4c}
        \end{subfigure}%
        \begin{subfigure}{0.20\textwidth}
                \centering
                \scal{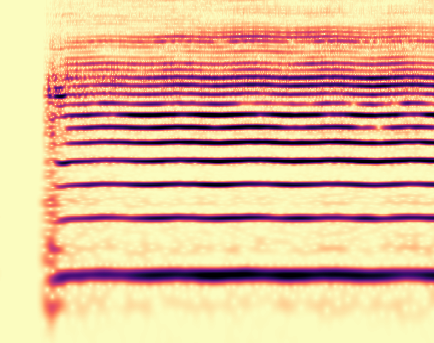}
                \caption*{4: \emph{ff}, on D string.}
                \label{fig:Vn-ord-G4-ff-3c}
        \end{subfigure}%

        \begin{subfigure}{0.20\textwidth}
                \centering
                \scal{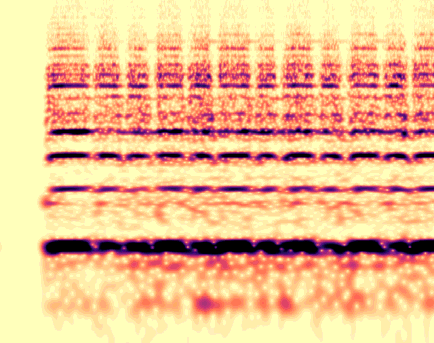}
                \caption*{5: \emph{tremolo}, C$\sharp$5, \emph{pp}.}
                \label{fig:Vn-trem-Csh5-pp-4c}
        \end{subfigure}%
        \begin{subfigure}{0.20\textwidth}
                \centering
                \scal{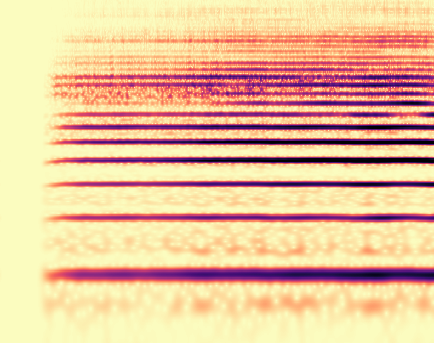}
                \caption*{5: on D string.}
                \label{fig:Vn-ord-G4-mf-3c}
        \end{subfigure}%

        \caption{Five nearest neighbors of the same query (a violin note with ordinary playing technique, at pitch G4, \emph{mf} dynamics, played on the G string), as retrieved by two different versions of our system: with MFCC features (left) and with scattering transform features (right). The captions denote the musical attribute(s) that differ from those of the query: mute, playing technique, pitch, and dynamics.
}\label{fig:demo-ordinary}
\end{figure}

\begin{figure}
        \newcommand{\scalwidth}{28.0mm}
        \newcommand{\scalheight}{22.1mm}
        \newcommand{\arrowextra}{2mm}

        \begin{subfigure}{\linewidth}
                \centering
                \scal{TpC-flatt-G4-mf.png}
                \caption*{Query: Trumpet in C, \emph{flatterzunge}, G4, \emph{mf}.}
                \label{fig:TpC-flatt-G4-mf}
        \end{subfigure}%

        \begin{subfigure}{0.20\textwidth}
                \centering
                \scal{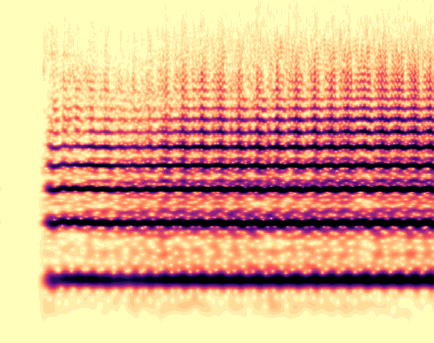}
                \caption*{1: F$\sharp$4.}
                \label{fig:TpC-flatt-Fsh4-mf-left}
        \end{subfigure}%
        \begin{subfigure}{0.20\textwidth}
                \centering
                \scal{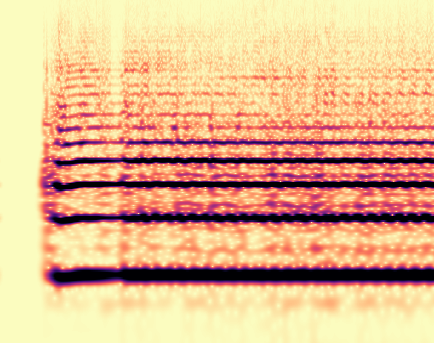}
                \caption*{1: \emph{pp}.}
                \label{fig:TpC-flatt-G4-pp}
        \end{subfigure}%

        \begin{subfigure}{0.20\textwidth}
                \centering
                \scal{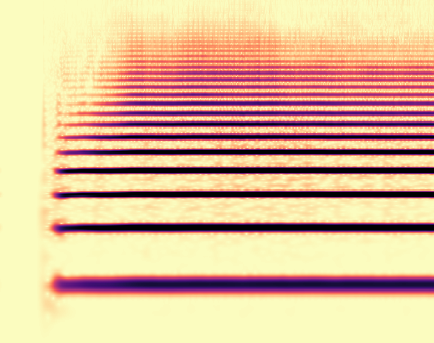}
                \caption*{2: \emph{ordinario}, F4.}
                \label{fig:TpC-ord-F4-mf}
        \end{subfigure}%
        \begin{subfigure}{0.20\textwidth}
                \centering
                \scal{TpC-flatt-Fsh4-mf.png}
                \caption*{2: F$\sharp$4.}
                \label{fig:TpC-flatt-Fsh4-mf-right}
        \end{subfigure}%

        \begin{subfigure}{0.20\textwidth}
                \centering
                \scal{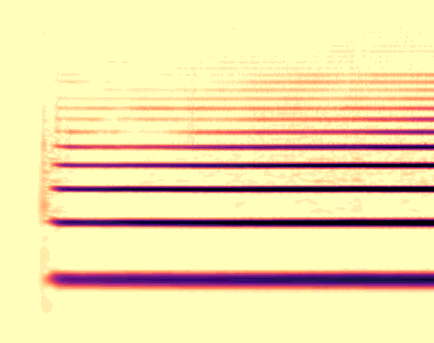}
                \caption*{3: \emph{ordinario}, F$\sharp$4.}
                \label{fig:TpC-ord-Fsh4-mf}
        \end{subfigure}%
        \begin{subfigure}{0.20\textwidth}
                \centering
                \scal{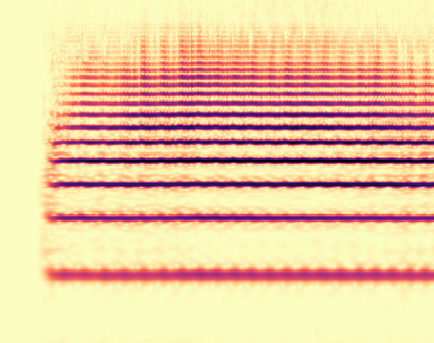}
                \caption*{3: straight mute.}
                \label{fig:TpC+S-flatt-G4-mf}
        \end{subfigure}%

        \begin{subfigure}{0.20\textwidth}
                \centering
                \scal{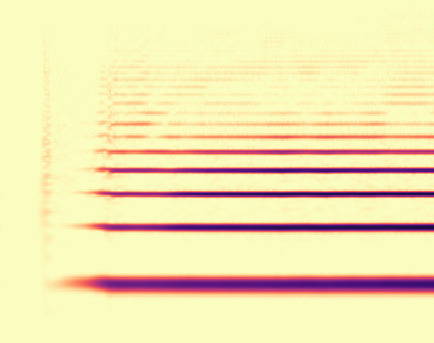}
                \caption*{4: \emph{ordinario}%
                , \emph{f}.}
                \label{fig:TpC-ord_flatt-F4-f}
        \end{subfigure}%
        \begin{subfigure}{0.20\textwidth}
                \centering
                \scal{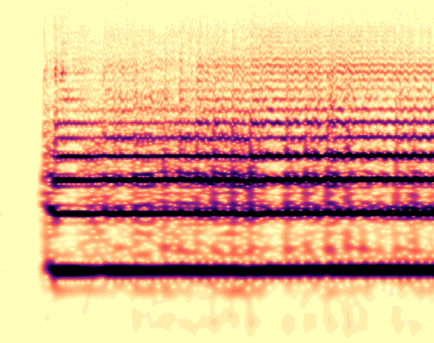}
                \caption*{4: G$\sharp$4, \emph{pp}.}
                \label{fig:TpC-flatt-Gsh4-pp}
        \end{subfigure}%

        \begin{subfigure}{0.20\textwidth}
                \centering
                \scal{TpC-ord-G4-mf.png}
                \caption*{5: \emph{ordinario}.}
                \label{fig:TpC-ord-G4-mf}
        \end{subfigure}%
        \begin{subfigure}{0.20\textwidth}
                \centering
                \scal{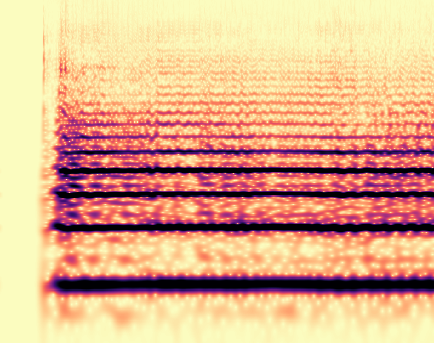}
                \caption*{5: F4, \emph{pp}.}
                \label{fig:TpC-flatt-F4-pp}
        \end{subfigure}%

        \caption{Five nearest neighbors of the same query (a trumpet note with \emph{flatterzunge} technique, at pitch G4, \emph{mf} dynamics), as retrieved by two different versions of our system: with MFCC features (left) and with scattering transform features (right).%
The captions of each subfigure denotes the musical attribute(s) that differ from those of the query.
}\label{fig:demo-extended}
\end{figure}

\subsection{Qualitative error analysis}
For demonstration purposes, we select an audio recording $\boldsymbol{x}(t)$ to query two versions of the proposed query-by-example system.
The first version uses MFCCs with $T=\SI{25}{\milli\second}$ and LMNN; it has a P@$5$ of $48.4\%$ for IPT retrieval.
The second version uses scattering coefficients with $T=\SI{1}{\second}$, logarithmic transformation with median renormalization (see Equation \ref{eq:log-scattering}), and LMNN; it has a P@$5$ of $63.0\%$ for IPT retrieval.
Both versions adopt IPT labels as reference for training LMNN.
The main difference between the two versions is the choice of spectrotemporal features.

Figure \ref{fig:demo-ordinary} shows the constant-$Q$ scalograms of the five retrieved items for both versions of the system as queried by the same audio signal $\boldsymbol{x}(t)$: a violin note from the SOL dataset, played with ordinary playing technique on the G string with pitch G4 and \emph{mf} dynamics.
Both versions correctly retrieve five violin notes which vary from the query in pitch, dynamics, string, and use of mute.
Therefore, both systems have an instrument retrieval P@$5$ of $100\%$ for this query.
However, although the scattering-based version is also $100\%$ correct in terms of IPT retrieval (\ie{}, it retrieves five \emph{ordinario} notes), the MFCC-based version is only $40\%$ correct.
Indeed, three recordings exhibit on of the \emph{tremolo} or \emph{sul ponticello} playing techniques.
We hypothesize that the confusion between \emph{ordinario} and \emph{tremolo} is caused by the presence of vibrato in the ordinary query since MFCCs cannot distinguish amplitude modulations (tremolo) from frequency modulations (vibrato) for the same modulation frequency \cite{anden2012dafx}.
These differences, however, are perceptually small and in some musical contexts vibrato and tremolo are used interchangeably.

The situation is different when querying both systems with recording $\boldsymbol{x}(t)$ exhibiting an extended rather than ordinary IPT.
Figure \ref{fig:demo-extended} is analogous to Figure \ref{fig:demo-ordinary} but with a different audio query.
The query is a trumpet note from the SOL dataset, played with the \emph{flatterzunge} (flutter-tonguing) technique, pitch G4, and \emph{mf} dynamics.
Again, the scattering-based version retrieves five recordings with the same instrument (trumpet) and IPT (\emph{flatterzunge}) as the query.
In contrast, four out of the five items retrieved by the MFCC system have an \emph{ordinario} IPT instead of \emph{flatterzunge}.
This shortcoming has direct implications on the usability of the MFCC query-by-example system for contemporary music creation.
More generally, this system is less reliable when queried with extended IPTs.

Unlike instrument similarity, IPT similarity seems to depend on long-range temporal dependencies in the audio signal.
In addition, it is not enough to capture the raw amplitude modulation provided by the second-order scattering coefficients.
Instead, an adaptive layer on top of this is needed to extract the discriminative elements from those coefficients.
Here, that layer consists of the LMNN metric learning algorithm, but other methods may work equally well.

\subsection{Feature space visualization}

To visualize the feature space generated by MFCCs and scattering transforms, we embed them using diffusion maps.
These embeddings preserve local distances while reducing dimensionality by forming a graph from those distances and calculating the eigenvectors of its graph Laplacian \cite{lafon2006acha}.
Diffusion maps have previously been used to successfully visualize scattering coefficients \cite{chudacek2014embc,villoutreix2017plos}.

\begin{figure*}
        \begin{subfigure}{0.45\textwidth}
                \centering
                \includegraphics[width=\linewidth]{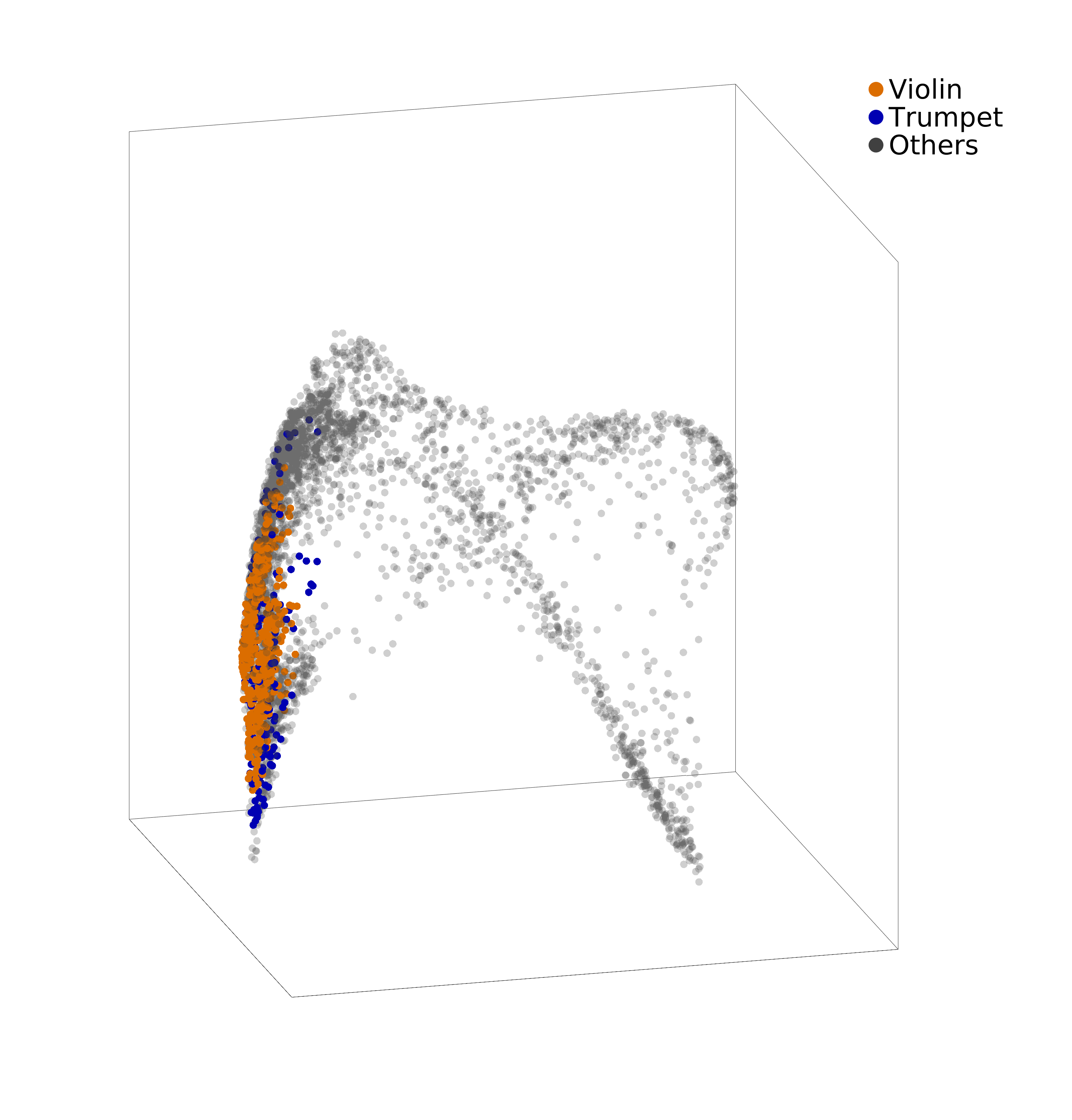}
                \caption{Instrument embedding with MFCC.}
                \label{fig:mf_inst_two_dmap}
        \end{subfigure}%
        \begin{subfigure}{0.45\textwidth}
                \centering
                \includegraphics[width=\linewidth]{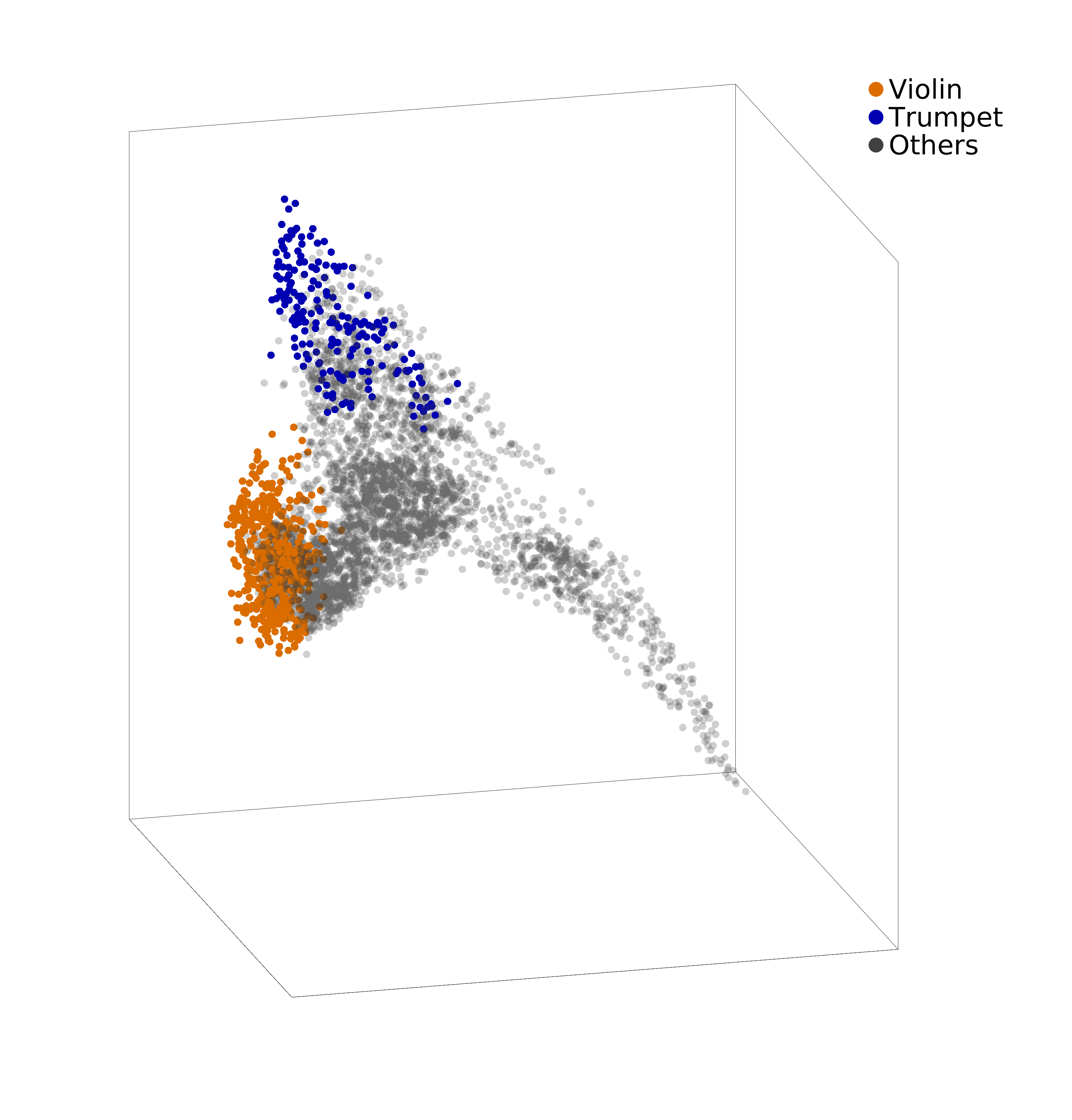}
                \caption{Instrument embedding with scattering transform.}
                \label{fig:sc_inst_two_dmap}
        \end{subfigure}%

        \begin{subfigure}{0.45\textwidth}
                \centering
                \includegraphics[width=\linewidth]{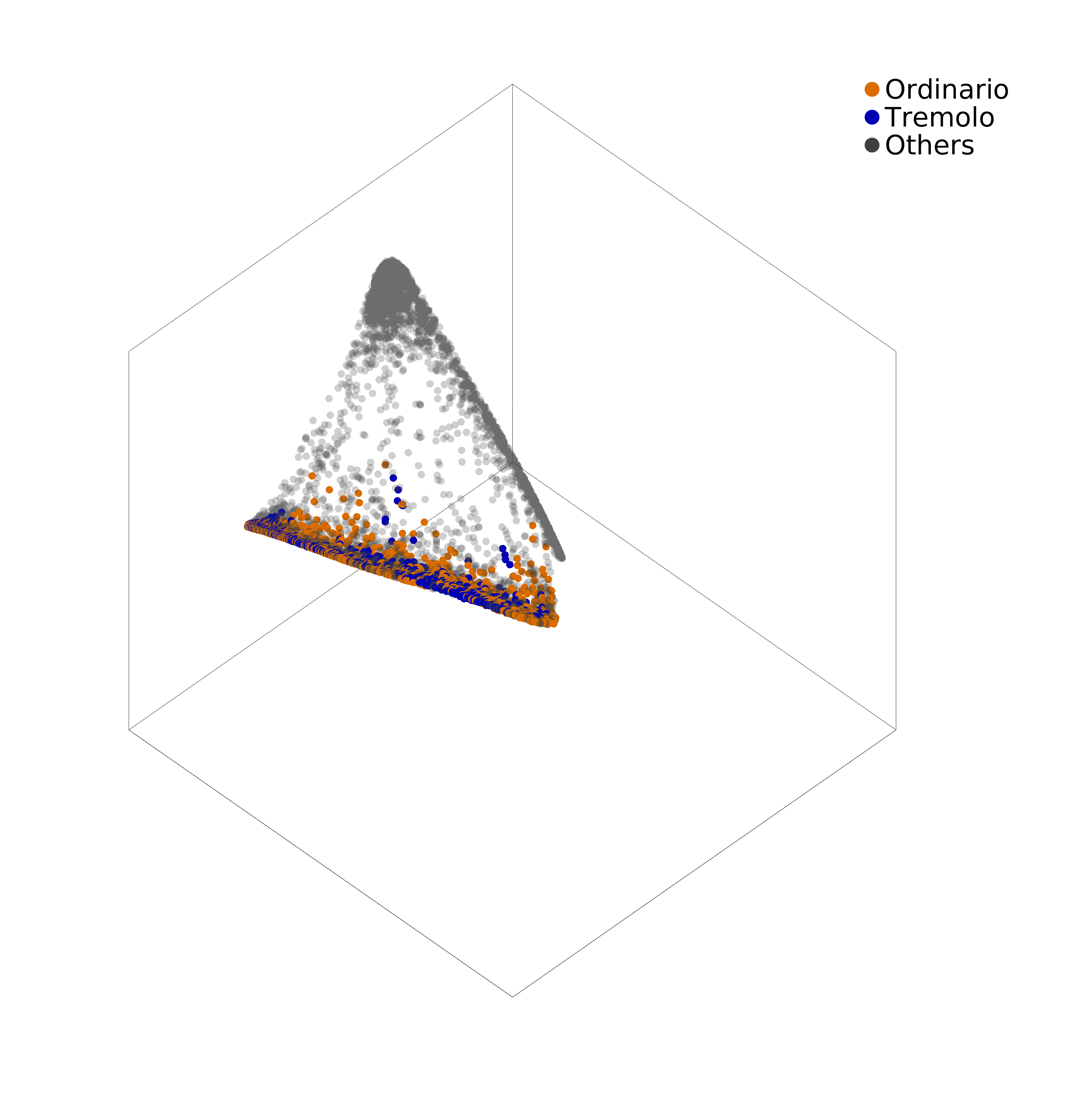}
                \caption{Playing technique embedding with MFCC.}
                \label{fig:mf_tech_two_dmap}
        \end{subfigure}%
        \begin{subfigure}{0.45\textwidth}
                \centering
                \includegraphics[width=\linewidth]{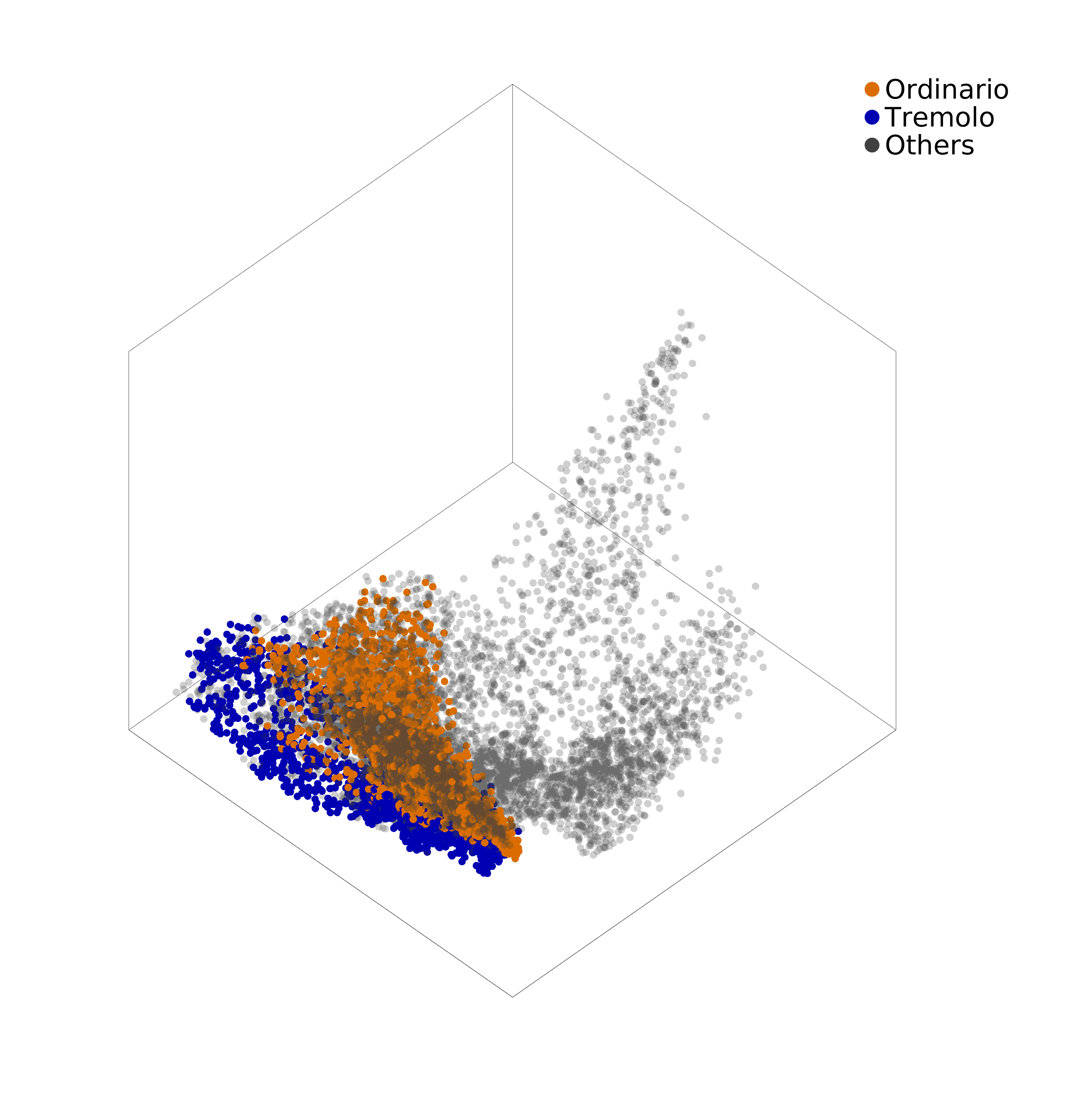}
                \caption{Playing technique embedding with scattering transform.}
                \label{fig:sc_tech_two_dmap}
        \end{subfigure}%

        \caption{Diffusion maps produce low-dimensional embeddings of MFCC features (left) \vs{} scattering transform features (right).
In the two top plots, each dot represents a different musical note, after restricting the SOL dataset to the \emph{ordinario} playing technique of each of the $31$ different instrument-mute couples. Blue (\resp{} orange) dots denote violin (\resp{} trumpet in C) notes, including notes played with a mute: \emph{sordina} and \emph{sordina piombo} (\resp{} \emph{cup}, \emph{harmon}, \emph{straight}, and \emph{wah}).
In the two bottom plots, each dot corresponds to a different musical note, after restricting the SOL dataset to $4$ bowed instruments (violin, viola, violoncello, and contrabass), and keeping all $38$ applicable techniques. Blue (\resp{} orange) dots denote tremolo (\resp{} ordinary) notes.
In both experiments, the time scales of both MFCC and scattering transform are set equal to $T=\SI{1}{s}$, and features are post-processed by means of the large-margin nearest neighbor (LMNN) metric learning algorithm, using playing technique labels as reference for reducing intra-class neighboring distances.}
        \label{fig:embeddings}
\end{figure*}

Figure \ref{fig:embeddings} shows embeddings of MFCCs and scattering coefficients, both post-processed using LMNN, for different subsets of recordings.
In Figure \ref{fig:mf_inst_two_dmap}, we see how the MFCCs fail to separate violin and trumpet notes for the \emph{ordinario} playing technique.
Scattering coefficients, on the other hand, successfully separate the instruments as seen in Figure \ref{fig:sc_inst_two_dmap}.
Similarly, Figures \ref{fig:mf_tech_two_dmap} and \ref{fig:sc_tech_two_dmap} show how, restricted to bowed instruments (violin, viola, violoncello, and contrabass), MFCCs do not separate the \emph{ordinario} from \emph{tremolo} playing techniques, while scattering coefficients discriminates well.
These visualizations provide motivation for our choice of scattering coefficients to represent single notes.

\section{Conclusion}

Whereas the MIR literature abounds on the topic of musical instrument recognition for so-called ``ordinary'' isolated notes and solo performances, little is known about the problem of retrieving the instrumental playing technique from an audio query within a fine-grained taxonomy.
Yet the knowledge of IPT is a precious source of musical information, not only to characterize the physical interaction between player and instrument, but also in the realm of contemporary music creation.
It also bears an interest for organizing digital libraries as a mid-level descriptor of musical style.
To the best of our knowledge, this paper is the first to benchmark query-by-example MIR systems according to a large-vocabulary, multi-instrument IPT reference ($143$ classes) instead of an instrument reference.
We find that this new task is considerably more challenging than musical instrument recognition as it amounts to characterizing spectrotemporal patterns at various scales and comparing them in a non-Euclidean way.
Although the combination of methods presented here -- wavelet scattering and large-margin nearest neighbors -- outperforms the MFCC baseline, its accuracy on the SOL dataset certainly leaves room for future improvements.
For example, we could replace the standard time scattering transform with
joint time-frequency scattering transform \cite{anden2018tsp}.

The evaluation methodology presented here uses ground truth IPT labels to quantify the relevance of returned items.
This approach is useful in that the labels are unambiguous, but it might be too coarse to reflect practical use.
Indeed, as it is often the case in MIR, some pairs of labels are subjectively more similar than others.
For example, \emph{slide} is evidently closer to \emph{glissando} than to \emph{pizzicato-bartok}.
The collection of subjective ratings for IPT similarity, and its comparison with automated ratings, is left as future work.
Another promising avenue of research is to formulate a structured prediction task for isolated musical notes, simultaneously estimating the pitch, dynamics, instrument, and IPT to construct a unified machine listening system, akin to a caption generator in computer vision.


\begin{acks}
The authors wish to thank Philippe Brandeis, \'{E}tienne Graindorge, St\'{e}phane Mallat, Adrien Mamou-Mani, and Yan Maresz for contributing to the TICEL research project
and Katherine Crocker for a helpful suggestion on the title of this article.
This work is supported by the ERC InvariantClass grant 320959.
\end{acks}

\clearpage

\bibliographystyle{ACM-Reference-Format}
\bibliography{sample-bibliography}

\end{document}